\documentclass[11pt,a4paper,twocolumn]{article}

\usepackage[utf8]{inputenc}
\usepackage[T1]{fontenc}
\usepackage{amsmath,amssymb,amsfonts}
\usepackage{graphicx}
\usepackage{natbib}
\usepackage{hyperref}
\usepackage{xcolor}
\usepackage{booktabs}
\usepackage{xspace}
\usepackage{subcaption}
\usepackage{stfloats}

\usepackage[activate={true,nocompatibility},final,tracking=true,kerning=true,spacing=true]{microtype}
\microtypesetup{expansion=false}

\usepackage[margin=1in]{geometry}
\usepackage{authblk}

\hypersetup{
	colorlinks=true,
	linkcolor=blue,
	filecolor=magenta,
	urlcolor=cyan,
	citecolor=blue,
	pdftitle={An Early Warning Model for Forced Displacement},
	pdfauthor={Geraldine Henningsen},
	pdfsubject={Anticipatory Action, Forced Displacement, Risk Modeling},
	pdfkeywords={Forced displacement, Anticipatory action, Risk modeling, Machine learning, Refugee flows},
}

\title{An Early Warning Model for Forced Displacement}

\author[1]{Geraldine Henningsen}
\affil[1]{Global Data Service, UNHCR, Marmorvej 51, 2100, Denmark. Email: hennings@unhcr.org}

\date{}  

\begin{document}
	
	\maketitle
	
	\begin{abstract}
		Monitoring tools for anticipatory action are increasingly gaining traction to improve the efficiency and timeliness of humanitarian responses. Whilst predictive models can now forecast conflicts with high accuracy, translating these predictions into potential forced displacement movements remains challenging because it is often unclear which precise events will trigger significant population movements. This paper presents a novel monitoring approach for refugee and asylum seeker flows that addresses this challenge. Using gradient boosting classification, we combine conflict forecasts with a comprehensive set of economic, political, and demographic variables to assess two distinct risks at the country of origin: the likelihood of significant displacement flows and the probability of sudden increases in these flows. The model generates country-specific monthly risk indices for these two events with prediction horizons of one, three, and six months.
		
		Our analysis shows high accuracy in predicting significant displacement flows and good accuracy in forecasting sudden increases in displacement--the latter being inherently more difficult to predict, given the complexity of displacement triggers. We achieve these results by including predictive factors beyond conflict, thereby demonstrating that forced displacement risks can be assessed through an integrated analysis of multiple country-level indicators. Whilst these risk indices provide valuable quantitative support for humanitarian planning, they should always be understood as decision-support tools within a broader analytical framework.
	\end{abstract}

\section{Introduction}    
\label{sec:intro}

Anticipatory action, or proactive planning and preparation for future crises, 
has the potential to significantly improve the efficiency and timeliness of humanitarian responses in times of crisis.
However, anticipatory action's success relies heavily on the ability to accurately detect and predict future crisis events, 
especially by reducing the risk of false alarms. 

Various early warning initiatives in the humanitarian sector have emerged in recent years to predict natural hazards and 
conflict by utilising quantitative modelling and machine learning \citep[e.g.,][]{Hegre2022, Mueller2022, Nevo2022}.
Whilst these initiatives have made significant progress in predicting natural hazards and conflicts, 
they have not been sufficient in predicting displacement movements within or across borders as
conflict and natural hazards' impact on forced displacement is heavily influenced by the environment in which they occur, 
making them unreliable as standalone predictors of forced displacement flows.

This paper addresses this gap by developing a novel monitoring approach that explicitly targets forced displacement risks.
We use gradient boosting classification to predict displacement risk and combine conflict forecasts with a comprehensive set of 
country-level indicators to predict two distinct risks: 
i., the likelihood of significant refugee flows and ii., the probability of sudden increases in these flows.
For both situations, our model provides monthly risk indices with one, three, and six-month prediction horizons, 
offering humanitarian organisations timely insights for operational planning.

Our approach advances existing work in three ways.
First, we move beyond single-factor predictors by integrating multiple variables influencing displacement decisions.
Second, we distinguish between ongoing significant displacement and sudden spikes in refugee flows, 
addressing different operational planning needs.
Third, we provide monthly updates with multiple prediction horizons, 
allowing organisations to balance short-term responses with longer-term preparation.

The remainder of this paper is structured as follows.
Section two gives a brief overview of the literature.
Section three describes our data sources and preparation methods.
Section four details the methodological approach, including our treatment of class imbalances and temporal dependencies.
Section five presents our results, analysing prediction accuracy and consistency over time.
We conclude with a discussion of practical applications and limitations, 
emphasising that whilst our risk indices provide valuable quantitative support for humanitarian planning, 
they should be understood as decision-support tools within a broader analytical framework.

\section{Literature overview}  
\label{sec:litrev}

In recent years, there has been a growing interest in quantitative approaches to predicting forced displacement.
However, most predictive work has focused on forecasting asylum seeker flows to specific regions, such as the European Union (EU).
\cite{Carammia2022} develop an adaptive machine learning algorithm to predict EU asylum applications up to four weeks ahead, 
using administrative statistics and non-traditional data sources like the Global Database of Events, Language, and Tone (GDELT) and Google Trends.
\cite{Boss2024} employ ensemble machine learning methods combining random forest and gradient boosting to forecast monthly asylum seeker flows to the EU27,
whilst \cite{Casagran2024} created a predictive tool for forecasting refugee and asylum seeker flows into Europe specifically for non-governmental organisations operating in European contexts.

Some studies have explored alternative methodological approaches.
\cite{Qi2023} develop a Flow-Specific Temporal Gravity model for forced displacement, 
demonstrating its superiority to traditional fixed-effects gravity models.
\cite{Susmann2024} propose a hierarchical Bayesian approach for projecting refugee and asylum seeker populations over longer time horizons.
Others have focused on specific displacement events, such as \cite{Wycoff2023},
who combine Twitter data with conflict event data to predict Ukrainian refugee flows to neighbouring countries.

Whilst these studies advance our understanding of forced displacement prediction and forecasting, 
they address only specific aspects of global displacement patterns.
Many focus on long-distance movements to Europe or the United States, 
representing only a fraction of global displacement movements, 
as most forcibly displaced people only flee to neighbouring countries.
Additionally, most existing approaches typically aim to predict exact displacement levels, 
which proves challenging in an operational and global context, 
given the complex and situational nature of forced displacement triggers and the uncertainty of displacement statistics, especially during crises.

Our work differs from previous studies in several aspects.
First, we develop a global model that assesses displacement risks at the country of origin level rather than focusing on specific destination regions.
Second, instead of predicting precise displacement levels like \cite{Susmann2024} or modelling specific displacement corridors like
\cite{Boss2024}, 
we estimate the risk of exceeding operationally relevant thresholds and the likelihood of sudden increases in displacement flows.
This classification approach is more reliable and practically useful in an early warning context than attempting to forecast exact displacement figures.
Third, our monthly predictions with one, three, and six-month horizons are specifically designed to support humanitarian operations, 
bridging the gap between academic modelling and practical application.

\section{Data}     
\label{sec:data}

\subsection{Dataset Structure and Coverage}
Our model is based on a longitudinal data set of country-month observations,
where a country is depicting the country of origin.
The data set comprises $n = 219$ countries and $t = 46$ months (from January 2020 to October 2024).
The longitudinal data set is balanced, which results in a total of 10,074 observations.

\subsection{Dependent Variables and Risk Classification}
The model's dependent variable aggregates monthly refugee and asylum seeker flows from the country of origin.
These flows have been derived from UNHCR's nowcasting project, which nowcasts monthly refugee and asylum seeker stocks and flows
using a combination of operational data, UNHCR's registration database, Eurostat, governmental statistics, and predictive modelling \citep{Panta2024}.
Our risk model identifies two types of movements in the flow data:

\begin{enumerate}
\item  Significant flows above certain yearly thresholds that are broken down to their monthly equivalents:
\begin{itemize}
\item 2,000 persons per year (baseline operational threshold)
\item 5,000 persons per year (medium operational threshold)
\item 10,000 persons per year (high operational threshold)
\item 25,000 persons per year (multi-agency response threshold)
\end{itemize}
\item Sudden increases in displacement counts when above these thresholds in both ongoing and new situations.
We only include sudden increases that exceed the respective threshold values.
\end{enumerate}

This dual focus on both absolute levels and sudden changes allows the model to capture different types of humanitarian challenges,
from gradual-onset crises to rapid deteriorations requiring an immediate response.

\subsection{Change Point Detection}

\begin{figure}[h!]
\centering
\includegraphics[width=0.8\linewidth]{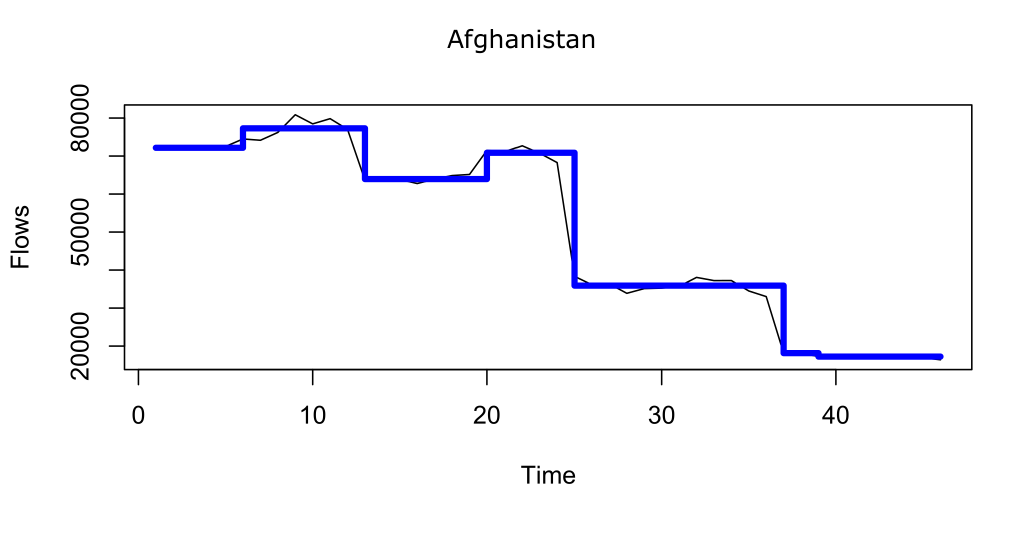}
\includegraphics[width=0.8\linewidth]{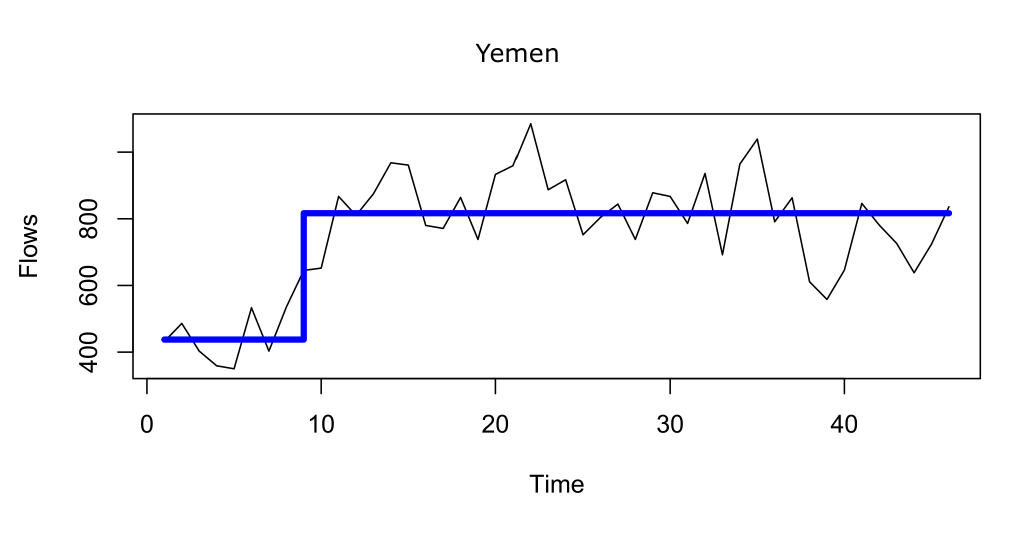}
\includegraphics[width=0.8\linewidth]{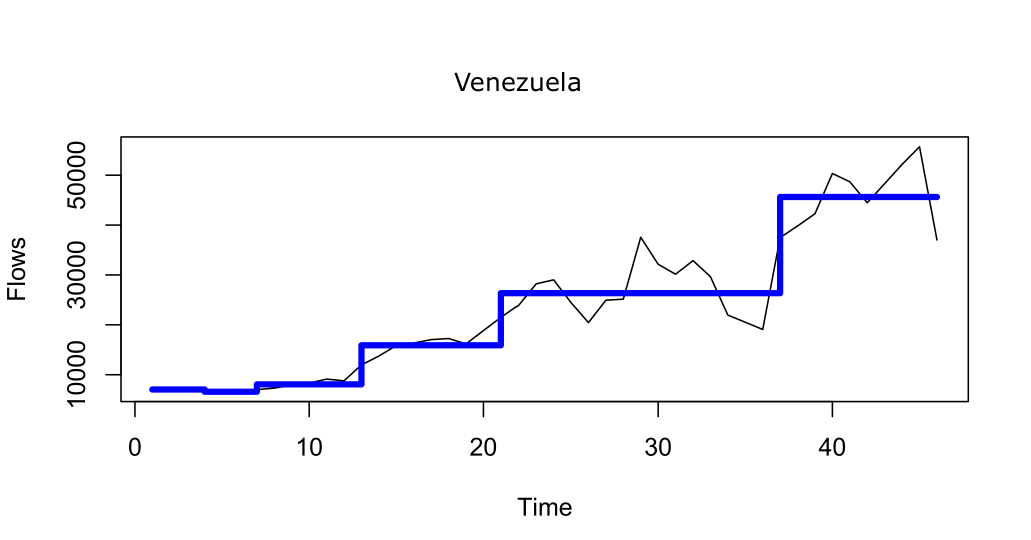}
\caption{Changepoints in the time series of refugee flows for Afghanistan, Yemen, and  Venezuela}
\label{fig:cp_ts}
\end{figure}

To identify sudden and significant changes in refugee and asylum seeker flows, we implement separate change point detection
models for each threshold level.
Following \cite{Carammia2022}, we use change point detection based on significant changes in the mean and
variance of the monthly time series.
We apply the `changepoint' package \cite{Killick2024} on all country individual
time series for refugees to identify months with a sudden change in
refugee flows (see examples for changepoints in the Afghanistan, Yemen, and Venezuela time series in Figure~\ref{fig:cp_ts}).

Combining thresholds and change points for each time series leads to six different scenarios (A--F) for a given month $t$,
as depicted in Table~\ref{tab:scen}.
\begin{table}[h!]
\begin{center}
\caption{Possible scenarios for flow time series at point $t$}
\label{tab:scen}
\begin{tabular}{l | c | c | c |}
& \multicolumn{3}{c}{Change point} \\
Threshold & Up & None & Down \\
\hline
Above &  A & B & C \\
\hline
Below &  D & E & F \\
\hline
\end{tabular}
\end{center}
\end{table}

However, from the perspective of an early warning model, only scenarios
A: `a sudden increase in the flow above the threshold value' and B and C: `a constant flow above the threshold value' are of interest.
This means that the remaining four scenarios can be grouped together, 
leading to three final classes for the dependent variable at a given month $t$:
\begin{itemize}
\item[]  \textit{Class 1}: Upwards changepoint above the threshold (scenario A)
\item[]  \textit{Class 2}: No upward changepoint but above the threshold (scenarios B and C)
\item[]  \textit{Class 3}: Every movement below the threshold (scenarios D, E, and F)
\end{itemize}

\begin{figure}[ht]\centering
	\includegraphics[width=0.8\linewidth]{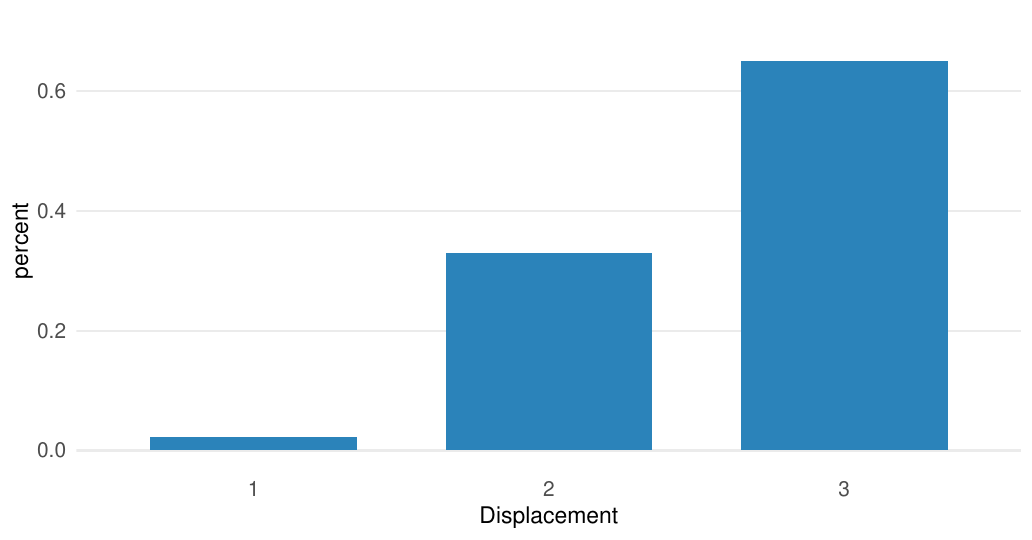}
	\caption{Proportion of observations in each class at threshold 2000}
	\label{fig:props}
\end{figure}

Categorising refugee/asylum seeker flows into these three classes leads,
unfortunately, to very unbalanced classes (see Figure~\ref{fig:props}).
An imbalance which increases with the threshold level.
As sudden and significant increases in the flow time series are a
rather seldom event, 
it is unsurprising that the proportion of observations that fall into class one is notably lower than the proportions of the other two classes.
We further discuss the impact of these significant class imbalances in the
methodology section.

\subsection{Feature variables}

We use 95 feature variables in our analysis (see table~2), covering conflict, economic, 
political, and demographic factors.
The core component consists of conflict predictions from Conflict Forecast \citep{Mueller2022},
a model targeted explicitly at detecting new or worsening conflicts.
We complement these conflict predictions with historical conflict data from ACLED \citep{Raleigh2010},
incorporating both event counts and fatalities by event category to capture conflict intensity and patterns.

Economic and market conditions are measured through indicators from \href{https://www.theglobaleconomy.com/download-data.php}{The Global Economy} and the World Bank, 
and food commodity prices from FAO's Food Price Monitoring Analysis (FPMA) database.
The latter focuses on staple foods, including bread, maize, potatoes, oil, rice, beans, and sorghum, as food insecurity often precedes displacement.

To capture political and social dynamics, we follow \citet{Carammia2022} and construct five indexes based on GDELT data \citep{Leetaru2013}:
Conflict, Economic Conditions, Political Stability, Social Cohesion, and Governance.
For each index, we calculate both the sum and the mean of the respective positive and negative Goldstein index across all events within each category, 
allowing us to capture the overall impact and average intensity of events.
We further include the average event tone to provide a comprehensive picture of the sociopolitical situation.

Natural hazard impacts are tracked using \href{https://www.emdat.be/}{EM-DAT} data, which provides information on event occurrence, financial damage, and affected populations.
Finally, we incorporate structural country characteristics through INFORM risk indices, including monthly severity assessments, coping capacity measures, 
vulnerability indicators, and hazard exposure metrics, supplemented by political, demographic, and geographic indicators from The Global Economy.

\begin{table*}[t]
	\caption{Overview of Feature Variables}
	\begin{tabular*}{\textwidth}{@{\extracolsep{\fill}}lll@{}}
		\toprule
		\textbf{Category} & \textbf{Description} & \textbf{Source} \\
		\midrule
		Conflict Prediction & Risk indicators for violence, intensity, armed conflict & Conflict Forecast \\
		Conflict Events & Event counts and fatalities by event type & ACLED \\
		Crisis Indicators & Severity, coping capacity, vulnerability indices & INFORM \\
		Economic & GDP, trade, industrial production, commodity prices & World Bank, FAO \\
		Socio-Political & Political stability, civil liberties, governance indices & Global Economy, GDELT \\
		Demographics & Population structure, urbanization, land use & Global Economy \\
		Natural Hazards & Deaths, injuries, damage, affected population & EM-DAT \\
		\bottomrule
	\end{tabular*}
\end{table*}

\section{Method}         
\label{sec:method}

\subsection{Data Preprocessing and Dimensionality Reduction}
Our preprocessing pipeline starts with the 95 feature variables described in Section~\ref{sec:data}.
Given the high dimensionality and potential collinearity of these features,
we employ Principal Component Analysis (PCA) for dimensionality reduction.
Before PCA, we standardize all continuous variables to have zero mean and unit variance.
The first five principal components are retained, capturing approximately 90\%
of the total variance in our feature space.
Notably, models using these five components consistently outperform those using the complete feature set,
suggesting that the PCA effectively captures the main signals and underlying patterns in the features
while reducing noise and convergence problems arising from the high dimensionality of the feature space.

\subsection{Model Architecture}
We implement a one-versus-one (OVO) approach using three separate gradient-boosting machines,
each addressing a binary classification problem between each pair of our three classes (sudden increase vs constant high flow, 
sudden increase vs no significant displacement, constant high flow vs no significant displacement).
This approach allows us to handle the inherent differences between class pairs better while maintaining model interpretability.

The temporal structure of our predictive models requires some consideration of the lags based on how different factors influence displacement.
Most feature variables, such as economic indicators or demographic factors, are slow-moving, 
and their effects on displacement typically manifest over time.
Therefore, these variables are lagged according to the respective prediction horizon to ensure
 we only use information that would be available at the time of prediction.
For example, in the 3-month prediction model, these variables are lagged by three months, 
in the 1-month prediction model by one month, and in the 6-month prediction by six.

However, conflict predictions require a different treatment because conflict often has an immediate impact on displacement decisions.
When violent events occur, populations may decide to flee quickly, 
making the temporal relationship between conflict and displacement more direct.
Therefore, for conflict prediction variables from Conflict Forecast, 
we use the predicted conflict risk corresponding to our target prediction month.
For instance, when predicting displacement risk for March 2024 using the 3-month horizon model (predicting in December 2023), 
we use Conflict Forecast's prediction for March 2024 rather than lagged values.
This alignment of conflict predictions with our target month better captures the immediate 
relationship between conflict events and displacement decisions.

Using conflict predictions of course adds an additional element of uncertainty to our risk index, 
as it will heavily depend on the accuracy of the conflict predictions. 
However, given the alternative of using real conflict data lagged by three or six months, 
we deem potential prediction errors as the lesser evil. 
This approach is theoretically justified by the temporal relationship between conflict and displacement---%
populations tend to react to imminent or developing conflict threats rather than historical patterns alone. 
Our analysis confirms this intuition, as models using predicted conflict values consistently outperform those relying solely on lagged conflict data. 
This suggests that despite introducing additional uncertainty, 
conflict predictions capture valuable forward-looking information that better aligns with the decision-making processes that drive displacement behaviors.

\subsection{Model Estimation and Hyperparameter Optimization}

We use the `h2o' package \cite{LeDell2022}, an R interface to H2O's distributed machine learning platform, to implement our gradient boosting machines.
Each model's hyperparameters are optimised through a random grid search over a comprehensive parameter space.
The search explores 150 random combinations from the following hyperparameter options, as outlined in table~3.
\begin{table*}[htbp]
	\caption{Hyperparameter Space for Gradient Boosting Models}
	\begin{center}
		\begin{tabular*}{0.7\textwidth}{@{\extracolsep{\fill}}ll@{}}
			\toprule
			\textbf{Hyperparameter} & \textbf{Range} \\
			\midrule
			Trees & \{200, 500, 1000\} \\
			Learning rate & \{0.001, 0.01, 0.1\} \\
			Max depth & \{3, 5, 9, 12\} \\
			Min rows & \{5, 10, 25\} \\
			Sample rate & \{0.8, 1.0\} \\
			Column rate & \{0.2, 0.5, 1.0\} \\
			\bottomrule
		\end{tabular*}
	\end{center}
\end{table*}
Due to significant class imbalances in our data, with sudden increases in displacement flows being relatively rare events,
we employ proportional stratified sampling for cross-validation.
This ensures that each fold maintains the same class distribution as the full dataset.
Due to the longitudinal structure of our data, 
we use moving time windows for our training and testing data to account for the temporal structure of the data.  
We use the Area Under the Precision-Recall Curve (AUCPR) as our primary metric for model selection,
as it is more informative than accuracy or area under the receiver operating characteristic curve (AUC-ROC) for imbalanced classification problems.

\subsection{Probability Calibration}

The OVO approach produces three separate binary classifiers, each comparing two classes while ignoring the third.
Therefore, we obtain three sets of binary probability estimates for each observation, one from each paired comparison.
To combine these into coherent three-class probabilities, we need to calibrate, rescale, and reconcile their predictions.

First, in order to create well-calibrated probabilities from the predicted values,
we use Platt scaling \citep{Platt1999} and use the test data as a framework to ensure that we use proportions that are as close as possible
to the class proportions of the predicted months.

Secondly, to make all three pairwise classification models comparable, we rescale the calibrated probabilities through
\begin{equation}
p_{i|jk}^{\text{rescaled}} = \Phi(\Phi^{-1}(p_{i|jk}^{\text{cal}}) - \Phi^{-1}(\text{threshold}_{jk}))
\end{equation}
Where:
\begin{itemize}
\item $p_{i|jk}^{\text{cal}}$ is the Platt scaled probability of class $i \in \{j, k\}$ from the classifier comparing classes $j$ and $k$
\item $p_{i|jk}^{\text{rescaled}}$ is the rescaled probability after transformation
\item $\Phi$ is the cumulative distribution function of the standard normal distribution
\item $\Phi^{-1}$ is the inverse cumulative distribution function (quantile function of the standard normal distribution)
\item $\text{threshold}_{jk}$ is the calibration threshold determined from the training data through precision-recall.
\end{itemize}

This transformation shifts the probability estimates in the normal quantile space to align the classification threshold with 0.5, ensuring 
that probabilities above 0.5 correspond to the desired class. After this rescaling, the complementary probability is calculated as:
\begin{equation}
p_{k|jk}^{\text{rescaled}} = 1 - p_{j|jk}^{\text{rescaled}}
\end{equation}

Finally, to derive consolidated probabilities $p_1, p_2, p_3$ that sum to unity and preserve the relative relationships from the binary models, 
we combine the rescaled probabilities. Let $p_{i|jk}$ now denote the rescaled probability for class $i \in \{j, k\}$ from the binary model comparing classes $j$ and $k$. 
Through a series of derivations considering pairwise probability ratios, we can express the final probabilities as:

\begin{equation}
p_i = \frac{1}{3} \sum_{(j,k) \in S_i} \frac{p_{i|ij} p_{i|ik}}{p_{i|ij} + p_{i|ik} - p_{i|ij}p_{i|ik}}
\end{equation}
Where:
\begin{itemize}
\item $S_i$ represents the set of pairs involving class $i$
\item $p_{i|ij}$ is the probability of class $i$ from the binary classifier comparing classes $i$ and $j$
\item $p_{i|ik}$ is the probability of class $i$ from the binary classifier comparing classes $i$ and $k$
\end{itemize}

This formulation ensures that: (i) all probabilities remain bounded between 0 and 1, 
(ii) the relative strengths of the binary predictions are preserved, 
and (iii) probabilities across all classes sum to unity. 
The final probability for each class is then calculated as the average of its reframed probabilities across all binary comparisons that contain said class.

\section{Results and Discussion}  
\label{sec:result}

We tested our model described in the previous section across multiple prediction horizons (1, 3, and 6 months)
and displacement thresholds (2,000, 5,000, 10,000, and 25,000 persons) for four respective time windows: March 2024 to June 2024.
The results remain relatively stable across different evaluation periods, indicating consistent model performance regardless of temporal context.
The following sections present detailed results for our evaluation period, focusing on performance metrics and predictive probabilities.

\subsection{Metrics}

Given the highly unbalanced class distribution in our data and the multiclass classification setting, 
we use several complementary metrics to evaluate model quality. 
To assess model performance, 
we compare the estimated probabilities produced by our model with the actual class labels derived from displacement data. 
Comparing probabilities with class labels allows us to evaluate not only the model's classification accuracy but also the quality and reliability of its probability estimates.

The log loss metric provides our primary measure of prediction accuracy, while the Brier score and Entropy score offer additional perspectives on model performance.
\begin{table*}[h!]
\caption{Performance metrics across different thresholds and prediction horizons}
\label{tab:metrics_overall}
\begin{center}
\begin{tabular}{lcccc}
\hline
Threshold & Horizon (months) & Log Loss & Brier Score & Entropy Score \\
\hline
2,000 & 1 & 0.26 & 0.12 & 0.36 \\
2,000 & 3 & 0.28 & 0.14 & 0.35 \\
2,000 & 6 & 0.31 & 0.16 & 0.49 \\
\hline
5,000 & 1 & 0.24 & 0.13 & 0.31 \\
5,000 & 3 & 0.24 & 0.12 & 0.32 \\
5,000 & 6 & 0.26 & 0.14 & 0.32 \\
\hline
10,000 & 1 & 0.19 & 0.10 & 0.28 \\
10,000 & 3 & 0.19 & 0.10 & 0.27 \\
10,000 & 6 & 0.20 & 0.10 & 0.30 \\
\hline
25,000 & 1 & 0.19 & 0.09 & 0.23 \\
25,000 & 3 & 0.18 & 0.09 & 0.21 \\
25,000 & 6 & 0.17 & 0.08 & 0.23 \\
\hline
\end{tabular}
\end{center}
\end{table*}

The metrics reveal a clear pattern in our model's predictive capabilities across different displacement magnitudes.
The consistent improvement in log loss, Brier score, and entropy metrics as thresholds increase suggests that larger
displacement events leave stronger and more detectable signals in our predictor variables.
This pattern makes intuitive sense from a humanitarian perspective---major displacement crises typically stem from more severe disruptions with clearer precursors,
while smaller displacement events may arise from more subtle or localized factors that are harder to detect systematically.
The lower entropy scores for higher thresholds confirm this,
indicating that the model can make more confident predictions about large-scale displacement events.
This capability to reliably identify major humanitarian crises several months in advance represents a significant operational advantage, 
as these events typically require the most substantial and coordinated response efforts.

\begin{table*}[h!]
\caption{Calibration metrics across different thresholds (1-month horizon)}
\label{tab:calibration_metrics}
\begin{center}
\begin{tabular}{lcccccc}
\hline
Threshold & ECE Class 1 & ECE Class 2 & ECE Class 3 & MCE Class 1 & MCE Class 2 & MCE Class 3 \\
\hline
2,000 & 0.10 & 0.13 & 0.03 & 0.27 & 0.45 & 0.45 \\
5,000 & 0.09 & 0.11 & 0.04 & 0.61 & 0.51 & 0.87 \\
10,000 & 0.07 & 0.08 & 0.03 & 0.44 & 0.45 & 0.43 \\
25,000 & 0.03 & 0.05 & 0.03 & 0.31 & 0.63 & 0.72 \\
\hline
\end{tabular}
\end{center}
\end{table*}

\begin{figure*}[h!]\centering
  \includegraphics[width=\linewidth]{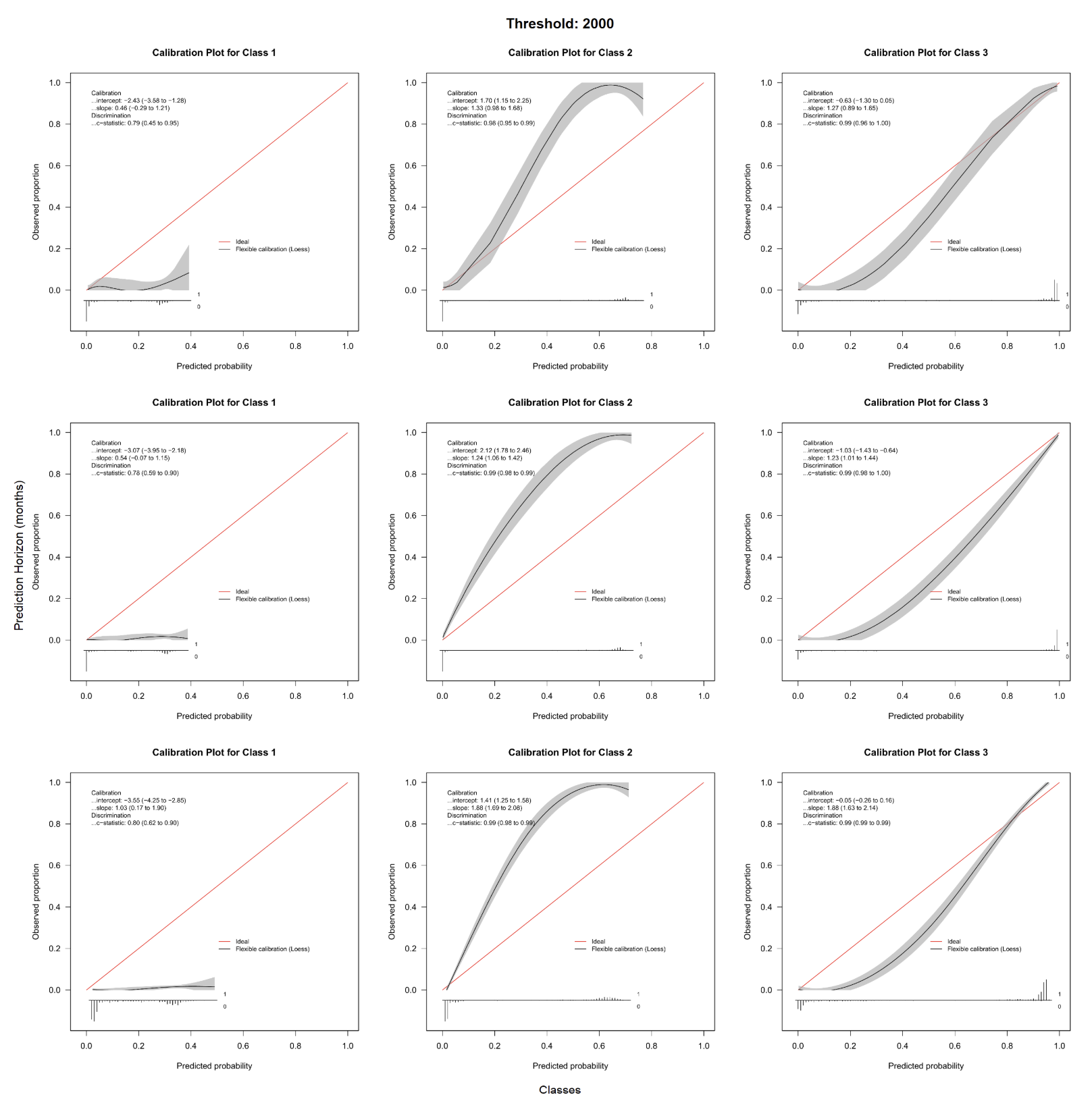}
  \caption{Calibration plots by class and predictive horizon for threshold 2000}
  \label{fig:cals}
\end{figure*}

Our calibration analysis reveals a pattern in the model's probability estimates.
While overall calibration appears strong (as shown by the overall low expected calibration error (ECE) values),
the significant gap between ECE and maximum calibration error (MCE) points to specific regions in the probability space where the model's estimates show higher uncertainty and require more careful interpretation.
The calibration plots in figure~\ref{fig:cals} indicate that the model tends to be overconfident in high-probability predictions for class~1 (sudden increases in displacement)
and underconfident in medium-range probabilities for class~2 (steady high displacement).
This characteristic of the probability estimates demonstrates the difficulty of finding a balance between capturing rare events like in class~1 and overall model performance.

Despite these calibration nuances, the overall metrics demonstrate that the model significantly
outperforms a simple benchmark that would only predict the majority class.
Given the class distribution of approximately 3:33:64 per cent across the three classes,
achieving discrimination capabilities consistently above 0.79 indicates strong predictive capability, 
as indicated by the concordance statistics (c-statistic) in Figure~\ref{fig:cals}.
It is also important to note that errors remain relatively balanced across the three classes,
indicating that the model does not systematically favour one class over others in its predictions.
This balanced error distribution is particularly valuable in an early warning context,
where false positives and negatives can have significant operational implications.

\begin{figure*}[h!]\centering
  \includegraphics[width=0.8\linewidth]{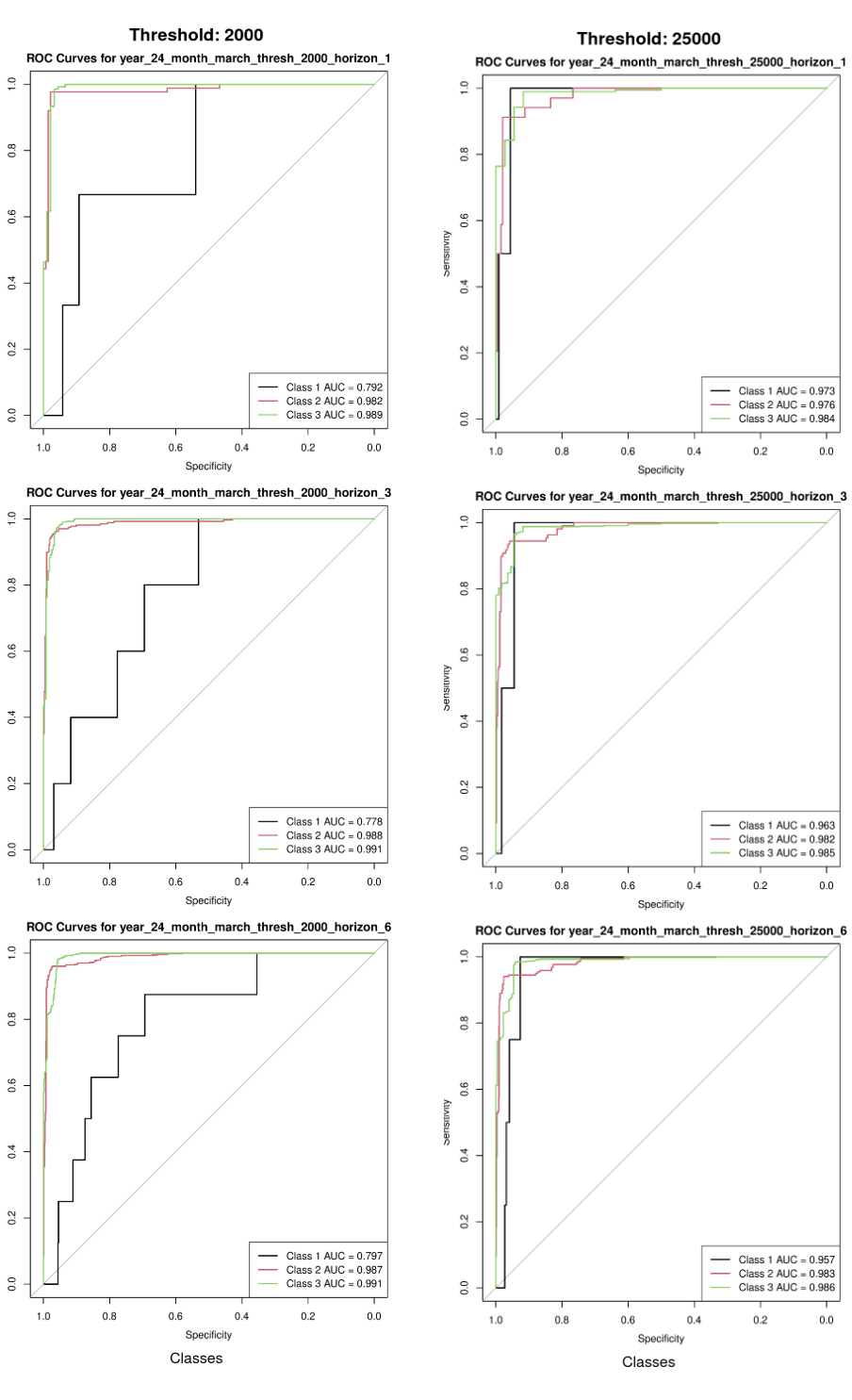}
  \caption{ROC curves for all three classes for each predictive horizon (1 month, 3 months, 6 months) - 2000 and 25000 threshold}
  \label{fig:rocs}
\end{figure*}

\begin{figure*}[h!]\centering
  \includegraphics[width=\linewidth]{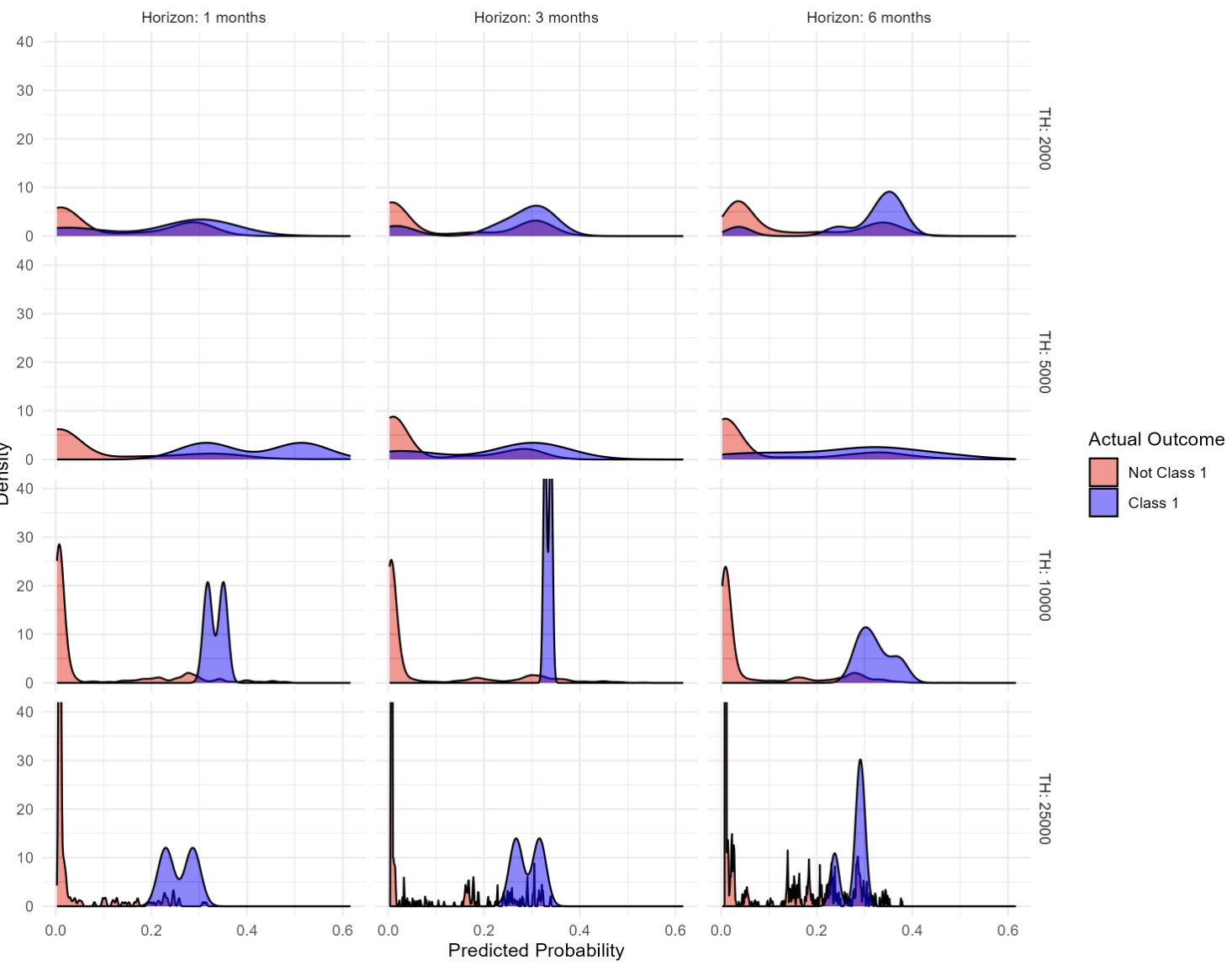}
  \caption{Kernel density plots class one against other classes by threshold and ptredictive horizon}
  \label{fig:kernel}
\end{figure*}

The receiver operating characteristic (ROC) curves in figure~\ref{fig:rocs} and density plots in figure~\ref{fig:kernel} further illustrate the model's strong discriminative ability,
particularly for higher displacement thresholds.
These visualizations demonstrate a clear separation between classes,
the kernel density plots show minimal overlap in predicted probabilities between class one events and non-class one events for the higher thresholds, 
albeit lower discrimination capabilities for lower thresholds.
This separation, especially for higher threshold levels, indicates that the model can effectively distinguish between different displacement scenarios,
providing reliable signals for different types of humanitarian planning needs.

\subsection{Predicted Probabilities}
\label{sec:probs}

The probability estimates produced by our model show distinct calibration patterns that provide insights into its predictive strengths and limitations.
While the model demonstrates good overall calibration (as indicated by low ECE values),
we observe more nuanced patterns when examining specific probability ranges.
Calibration diagrams (see Figure~\ref{fig:cals}) reveal that the model tends to be overconfident in high-probability predictions for class one (sudden increases in displacement),
particularly in the 0.3-0.5 probability range.
Conversely, for class 2 (constant high displacement),
the model shows a tendency toward underconfidence in medium probability ranges.

These calibration patterns remain consistent across different displacement thresholds,
though they appear most pronounced at lower thresholds (2,000-5,000 displaced persons).
At higher thresholds, calibration improves overall,
with only minor deviations between predicted probabilities and observed frequencies.
This pattern suggests that the model's probability estimates become more reliable as the magnitude of displacement increases,
which aligns with our earlier findings on overall performance metrics.

\begin{figure*}[h!]
	\centering
	\begin{subfigure}[b]{0.85\textwidth}
		\centering
		\includegraphics[width=0.65\textwidth]{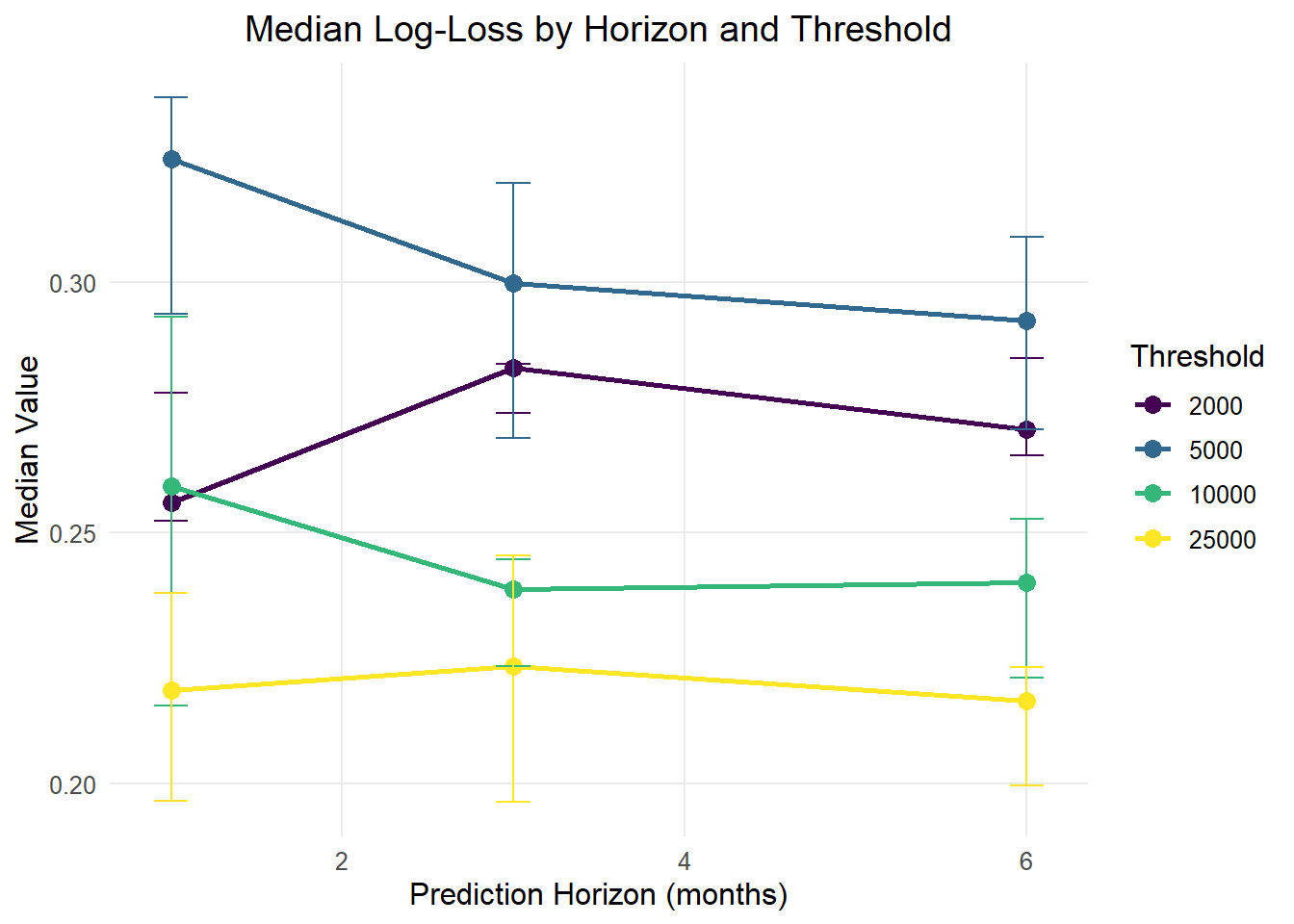}
		\caption{Log Loss by horizon and threshold}
		\label{fig:log_loss}
	\end{subfigure}
	
	\vspace{1cm}
	
	\begin{subfigure}[b]{0.85\textwidth}
		\centering
		\includegraphics[width=0.65\textwidth]{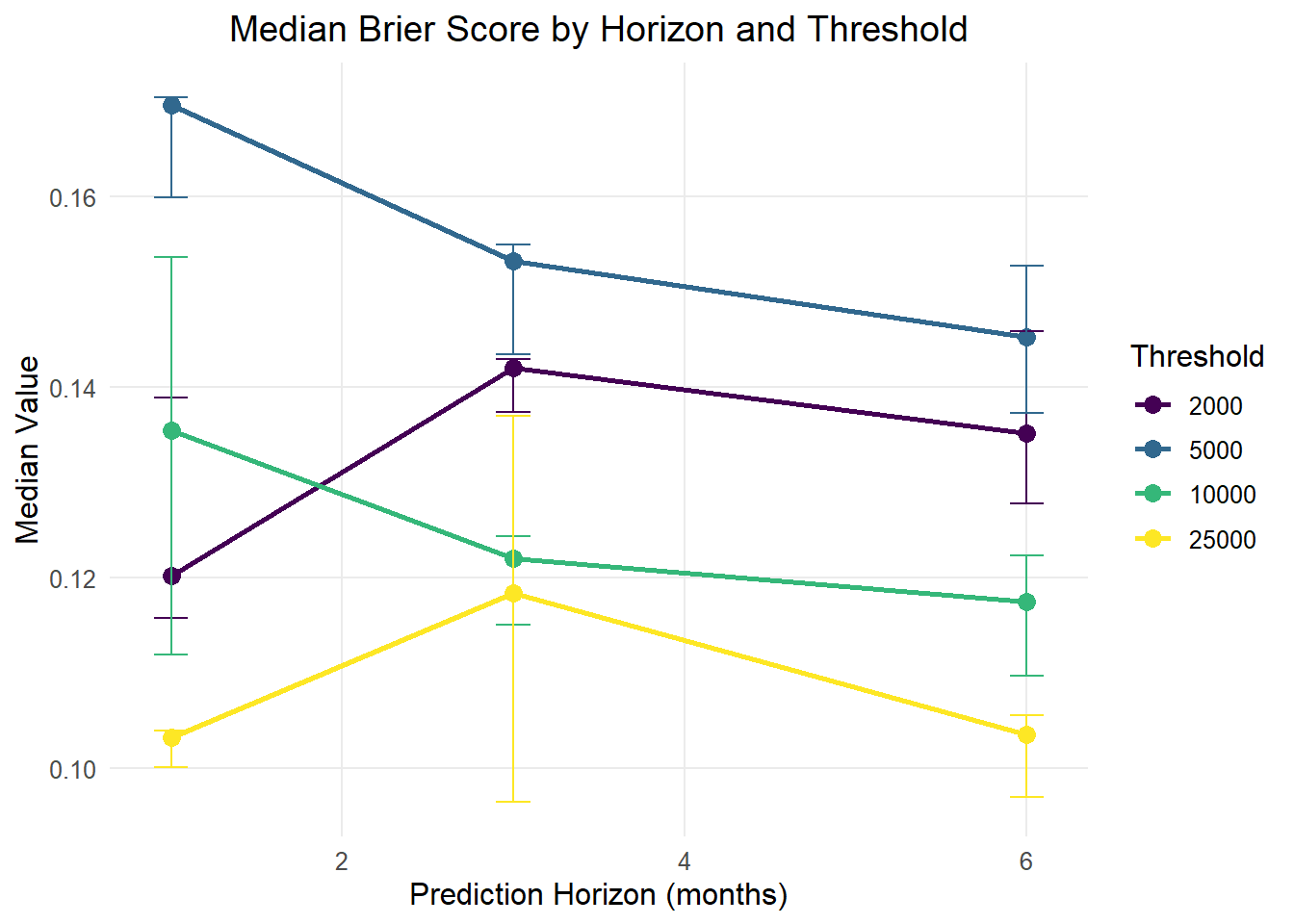}
		\caption{Brier Score by horizon and threshold}
		\label{fig:brier}
	\end{subfigure}
	
	\vspace{1cm}
	
	\begin{subfigure}[b]{0.85\textwidth}
		\centering
		\includegraphics[width=0.65\textwidth]{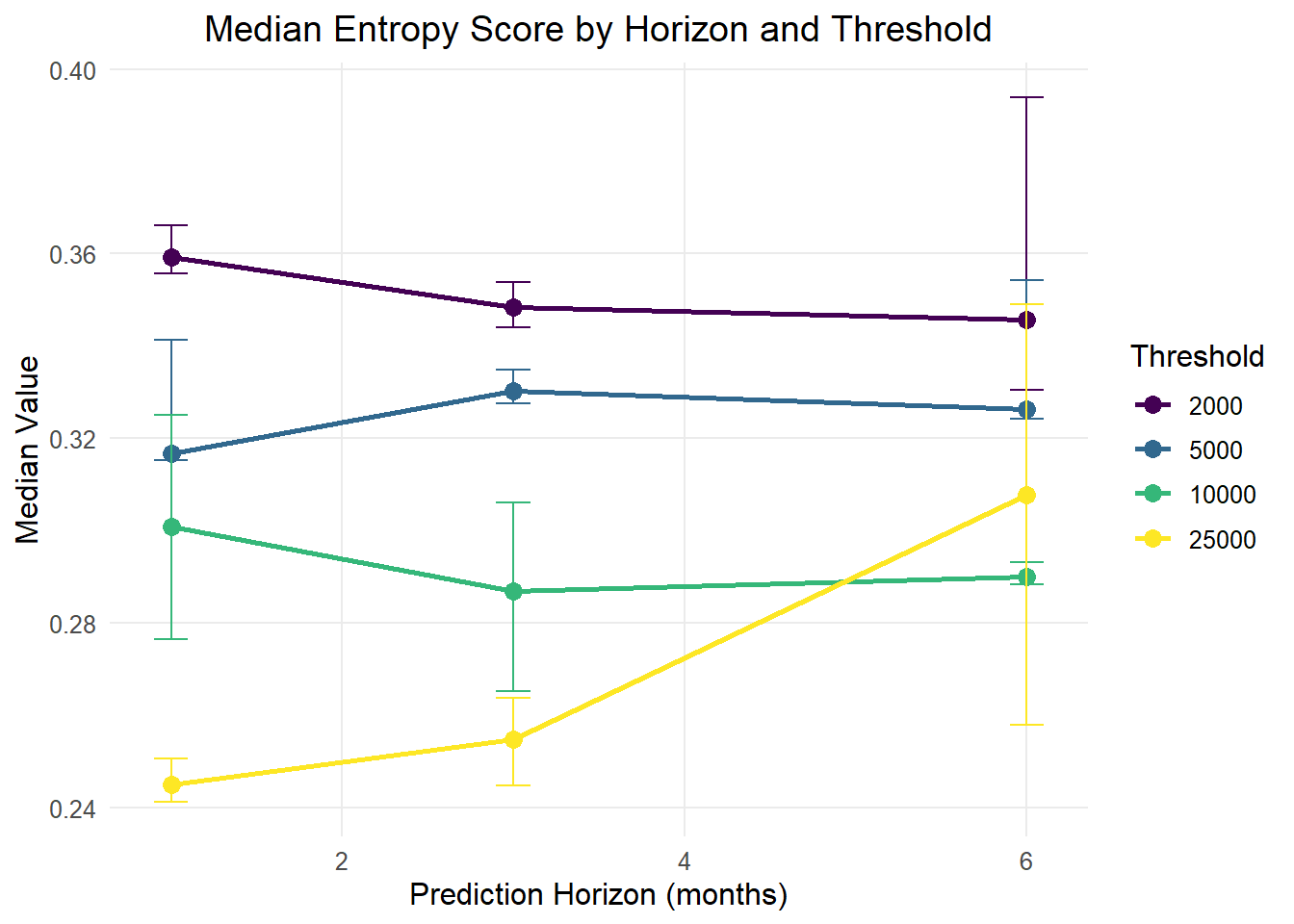}
		\caption{Entropy Score by horizon and threshold}
		\label{fig:entropy}
	\end{subfigure}
	\caption{Model performance metrics across different prediction horizons and displacement thresholds}
	\label{fig:metrics_horizon}
\end{figure*}

Our model demonstrates high temporal stability, as illustrated in Figure~\ref{fig:metrics_horizon},
which displays the mean log loss, Brier score, and entropy metrics and their spread across the four time windows in our sample
for different prediction horizons and displacement thresholds.
These charts reveal that average performance metrics show minimal degradation as prediction horizons extend from one to six months.
The stability varies by displacement threshold, while the 2,000- and 5,000-person thresholds (blue and purple lines)
show more spread across time windows with wider confidence intervals,
the 10,000 and 25,000 person thresholds (green and yellow lines) maintain consistently strong performance with narrower confidence intervals.
This pattern supports our earlier findings that larger displacement events generate stronger signals that persist over time,
allowing the model to maintain predictive power even at extended horizons.
The decreasing variance across time windows for higher thresholds indicates more reliable predictions for major displacement events
regardless of when the prediction is made.
This temporal consistency provides humanitarian organizations with dependable early warnings up to six months in advance—%
critical lead time for resource mobilization and operational planning before displacement crises materialize.

Despite the model's strong performance, several limitations must be acknowledged when applying it in humanitarian contexts.
First, the model's calibration shows some overconfidence in high-probability predictions for sudden displacement increases,
potentially leading to resource misallocation if taken at face value.
Second, while accuracy improves for larger displacement events,
the model is less reliable for smaller movements below 5,000 persons,
which still represent significant humanitarian challenges, 
particularly in resource-constrained environments.
Third, the model depends heavily on the quality of conflict forecasts,
which compounds uncertainty when these predictions are inaccurate.

From an operational perspective, these limitations suggest that the model should serve as a decision-support tool rather than an
automated trigger for humanitarian action.
Organizations should integrate these quantitative risk indices within a broader analytical framework that incorporates qualitative assessment,
local expertise, and contextual understanding.
Furthermore, the model should be regularly recalibrated as new data becomes available,
particularly following major crises that may reveal new displacement patterns not yet captured in historical data.
When properly contextualized within these constraints,
the model can significantly enhance anticipatory approaches to forced displacement without supplanting human judgment in humanitarian decision-making.

\section{Conclusion}   
\label{sec:conclude}
This study introduces a novel monitoring approach for anticipating forced displacement flows using gradient boosting classification. 
By combining conflict forecasts with economic, political, and demographic variables, 
we developed a model that assesses two critical risks: the likelihood of significant displacement flows and the probability of sudden increases in these flows. 
The model provides country-specific monthly risk indices with prediction horizons of one-, three-, and six months, 
offering humanitarian organizations valuable lead time for operational planning.

Our results demonstrate that the model accurately predicts significant displacement events, particularly for higher displacement thresholds. 
This finding suggests that larger-scale displacement crises leave stronger, more detectable signals in our predictor variables than smaller movements. 
The relative stability of performance metrics across different prediction horizons indicates that these signals persist over time, 
enabling reliable early warnings even at longer time frames.

Future research could extend this approach through a more granular geographic analysis and by incorporating additional data sources, 
such as satellite imagery. 
As anticipatory approaches to humanitarian action continue to evolve, 
quantitative models like ours can provide valuable decision support while complementing qualitative analysis and local expertise.
Ultimately, while no model can perfectly predict human behaviour in crises, 
our approach provides a systematic framework for anticipating forced displacement risks. 
This contribution offers humanitarian organizations a practical tool for more proactive planning and resource allocation.

\bibliographystyle{apalike}
\bibliography{references}  

\end{document}